\documentclass[aps,preprint,showpacs,superscriptaddress,nofootinbib]{revtex4-1}%
\usepackage{mathrsfs}
\usepackage{amsfonts}
\usepackage{amsmath}
\usepackage{amssymb}
\usepackage{graphicx}
\begin{document}
\title{Two-way deterministic quantum key distribution against detector side channel attacks}
\author{Hua Lu}
\email{hua.lue@gmail.com}
\affiliation{Department of Mathematics and Physics, Hubei University of Technology, Wuhan
430068, China}
\affiliation{State Key Laboratory of Magnetic Resonances and Atomic and Molecular Physics,
Wuhan Institute of Physics and Mathematics, Chinese Academy of Sciences, Wuhan 430071, China}
\author{Chi-Hang Fred Fung}
\email{chffung@hku.hk}
\affiliation{Department of Physics and Center of Theoretical and Computational Physics,
University of Hong Kong, Pokfulam Road, Hong Kong}
\author{Qing-yu Cai}
\email{qycai@wipm.ac.cn}
\affiliation{State Key Laboratory of Magnetic Resonances and Atomic and Molecular Physics,
Wuhan Institute of Physics and Mathematics, Chinese Academy of Sciences, Wuhan 430071, China}

\begin{abstract}

In a two-way deterministic quantum key distribution (DQKD) protocol, Bob randomly prepares
qubits in one of four states and sends them to Alice.
To encode a bit, Alice performs an operation on each received qubit and returns it to Bob.
Bob then measures the backward qubits to
learn about Alice's operations and hence the key bits.
Recently, we proved the unconditional security of the final key of this protocol in the ideal device setting.
In this paper, we prove that two-way DQKD protocols are immune to
all detector side channel attacks at Bob's side,
while we assume ideal detectors at Alice's side for error testing.
Our result represents a step forward in making DQKD protocols secure against general detector side channel attacks.

\end{abstract}


\maketitle

\section{introduction}

Quantum cryptography, also called quantum key distribution (QKD), allows two
remote parties, usually called Alice and Bob, to establish a secret
key that can be used to transmit secret message with a classical one-time pad
scheme. After the pioneering work presented by Bennett and Brassard in 1984
(BB84), the unconditional security of the BB84 protocol was proven with the
assumption of ideal
settings \cite{bb84,mayer01, lochau99, sp00}. In practice, the devices of QKD
are imperfect, the eavesdropper, called Eve, can exploit the imperfections of
the devices to gain partial or full information about the key bits
\cite{gllp,gisin06}. It has been shown that Eve may take advantage of the
nonzero multiphoton emission probability of Alice's laser source and the loss
of quantum channel, by using the photon-number-splitting (PNS) attack
\cite{bbb92, fgsz01, glms00, lutken00}. Decoy state QKD has been proposed to
beat Eve's PNS attacks, which can substantially increase the secure key
distribution distance \cite{hwang03,wang05,lmc05}.

An important part of QKD system is the measurement devices. In experiments, the
detection efficiency of the practical single-photon detectors (SPDs) is low
and in most cases, the two detectors are asymmetric, which may be exploited by
the Eve to steal information on the final key. For example,
attacks that exploit the efficiency mismatch between two SPDs in QKD system have been proposed,
demonstrated, and analyzed; they include the time-shift attack~\cite{qflm06,zfqcl07,Fung:Mismatch:2009},
and the faked states attack~\cite{mas05}.
Also, it has been
demonstrated experimentally that Eve can tracelessly acquire the full secret
key bits using specially tailored bright illumination \cite{vm10}.
Recently, a measurement-device-independent (MDI) QKD scheme was proposed to remove
detector side channel attacks \cite{lcq12}.
The advantage is that security can be guaranteed without regard to the experimental details of the measurement device.
In this scheme, Alice and Bob send signal pulses
to a Bell-state measurement device which may be owned by Eve and it is proven that they can
distill some secure key bits based on the Bell measurement results.

Recently, the security of four-state
deterministic quantum key distribution (DQKD) with a two-way quantum channel against the most
general attacks has been proven by us \cite{lfmc11,fmcc12} under the ideal-device setting,
while it is widely believed that two-way DQKD is vulnerable under Eve's practical attacks
because Eve can attack the travel photon in both forward and backward lines.
In this paper, we prove that
two variations of the DQKD protocol are immune to
all detector side channel attacks at Bob's side in the backward line.
We do not analyze the MDI security of
Alice's detectors and assume that they are ideal and non-blinded.

\section{security of the four-state protocol}
\label{sec-security of the four-state protocol}

In most QKD protocols,
after distributing all signal qubits, Alice and Bob should communicate
through public channel for basis reconciliation. In the BB84 protocol, about half of Bob's
measurement results are discarded due to the inconsistent bases Alice and Bob used
\footnote{In the efficient BB84 protocol~\cite{EffBB84_05}, the fraction of the
measurement results to be discarded goes to zero asymptotically.}.
But, in two-way DQKD protocols
\cite{bf02,cai03,cai04pra,cai04,deng04,lm05}, Bob can decode Alice's key bits
after measuring the returned qubits directly, without basis reconciliation.
This means that all measurement results will be used for key distillation
\footnote{This is true for any number of signals for the two-way protocol, not just asymptotically.}.
The two-way DQKD was first proposed by using entanglement \cite{bf02}.
Later, single-photon two-state protocol \cite{cai04} was proposed for
improving the experimental performance and then single-photon four-state protocol
was proposed independently in Refs. \cite{deng04,lm05}.
After that, the security of two-way DQKD against some special attacks were studied
\cite{cai03,cai06}. Recently, the security of the four-state protocol
against the most general attacks under the ideal-device setting was proved by us \cite{lfmc11,fmcc12}.

\begin{figure}
\begin{center}
\includegraphics[width=5in]{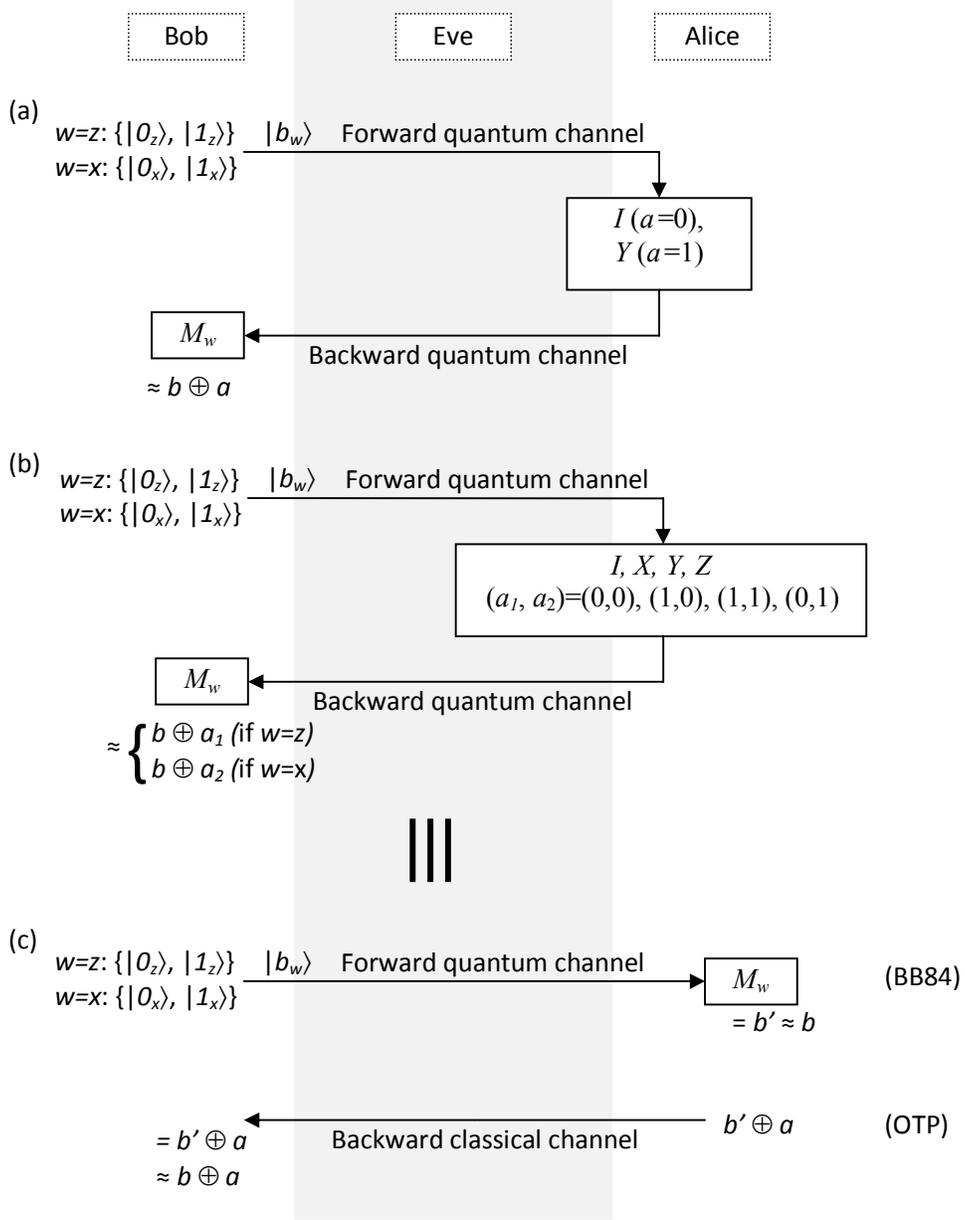}
\caption{
\label{fig:protocols}
We prove that the four-state DQKD protocols with two encoding operations (a) and four encoding operations (b)
are MDI secure in line A-to-B.
Protocol (b) was shown to be equivalent to protocol (c)
where the forward channel runs the BB84 protocol
and the backward channel runs a classical one-time pad (OTP)~\cite{fmcc12}.
Here, $a$, $a_1$, and $a_2$ represent a raw secret key bit in all three protocols randomly chosen by Alice.
In protocol (b), either $a_1$ or $a_2$ will be used as the key bit depending on the basis chosen by Bob
which he will inform Alice after his measurement.
Here, $|b_w\rangle$ represents eigenstate $b$ of basis $w$ and $M_w$ represents a measurement in basis $w$.
}
\end{center}
\end{figure}

Let us start with a brief review of the four-state two-way DQKD
protocol with two encoding operations [see Fig.~\ref{fig:protocols}(a)]. (1) Bob
prepares $n$ qubits randomly in one of the four states, $|0\rangle$,
$|1\rangle$, $|+\rangle$, and $|-\rangle$, where $|\pm\rangle= (|0\rangle
\pm|1\rangle)\sqrt{2}$ and sends them to Alice. (2) Alice randomly switches
the communication to the check mode or the encoding mode. (3) In check mode,
Alice randomly measures part of the received states in the $X$ basis or $Z$
basis. (4) In the encoding mode, Alice randomly perform unitary operations
$I=|0\rangle\langle0|+|1\rangle\langle1|$ to encode bit 0 or $Y=|0\rangle
\langle1|-|1\rangle\langle0|$ to encode bit 1. (5) Alice sends the encoded
qubits back to Bob. Since Alice's encoding operations do not change the bases
of Bob's states, \textit{i.e.}, $Y\{|0\rangle,|1\rangle\}=\{-|1\rangle
,|0\rangle\}$ and $Y\{|+\rangle,|-\rangle\}=\{|-\rangle,-|+\rangle\}$, Bob
measures each qubit in the same basis as the one he used for preparation to
get Alice's key bit deterministically, without basis reconciliation. (6) After
Bob measured all returned qubits, Alice announces her measurement results in the check mode
to compute the fidelity of the forward states in the B-to-A channel for the
consistent-basis measurements. They can get the fidelity $f_{0}$, $f_{1}$,
$f_{+}$, and $f_{-}$ of $|0\rangle$, $|1\rangle$, $|+\rangle$, and $|-\rangle
$, respectively. (7) Alice announces part of her key bits in encoding mode
to compute the error-rate $e$ in the A-to-B channel. (8) Alice and Bob perform
error-correction (EC) and privacy amplification (PA) to generate the final key
bits. In the asymptotic scenario, after verifying $\xi\equiv(f_{0}+f_{1}+f_{+}
+f_{-})/2-1\ge1/2$, Alice and Bob can get the secure final key against general
attacks with the generation rate,
\begin{equation}
\label{key:rate}
r=1-h(\xi)-h(e),
\end{equation}
where $h(x)=-x \log x - (1-x) \log (1-x)$ is the binary entropy function, and
$h(\xi)$ and $h(e)$ are the amounts of key bits Alice and Bob should
sacrifice for EC and PA.

With the idea of delaying PA, the security of a two-way DQKD protocol that uses four
states and four encoding operations was proven \cite{fmcc12} [see Fig.~\ref{fig:protocols}(b)].
It has been shown that the idea of delaying PA can simplify the security proof of two-way
DQKD protocols, with the same key generation rate as that derived in
\cite{lfmc11}. The security of two-way DQKD can also be proved against most
general attacks with the entropic uncertainty relation, and a higher final key
generation rate may be achieved \cite{blmr13}.

\section{Security of the four-state protocol against all detector side channel
attacks in line A-to-B}

We prove in this section the MDI security of line A-to-B for the original
four-state protocol with two encoding operations proposed by Refs. \cite{deng04,lm05},
and the four-state protocol with four encoding operations proposed by Ref. \cite{fmcc12}.
Our proof relies on the use of equivalent protocols.

\subsection{Proof of MDI security of the four-state protocol with two encoding operations}
\label{sec-proof-MDI1}
In order to prove the security of the MDI version of the protocol in line A-to-B,
we modify the protocol by moving Bob's measurement to the hands of Eve in line A-to-B.
The modified protocol is the same as the original protocol in
Sec.~\ref{sec-security of the four-state protocol} except with step (5) replaced as follows:
(5) Alice sends the encoded qubit to Eve.
Bob tells Eve the measurement basis.
Eve makes a measurement on Alice's qubit using Bob's basis, and returns the measurement result to Bob.

Note that Bob tells Eve the basis only after the Alice has received the qubit on line B-to-A.
This ensures that Eve cannot determine the initial qubit Bob sent out.
Intuitively, we can understand the security of the modified protocol as follows.
Since Bob randomly prepares the signal qubit in one of the four states, Eve cannot
know Alice's encoding operations even if she gets Bob's measurement result.
To know Alice's encoding, Eve has to know Bob's initial state as well.

To prove security of the modified protocol rigorously, we consider the PA term in the key rate formula.
The secure key
rate $r_{PA}$ from PA for secret key generation is given by
Renner and K\"onig's result~\cite{rk05}:
$r_{PA} =
S(\rho^{A}|\rho^{BE}, \text{basis}) = S(\rho^{ABE}| \text{basis})-S(\rho^{BE}| \text{basis})$,
where $\rho^{ABE}$ is the overall state of Alice's encoding ($A$), the qubit emitted
to line B-to-A ($B$), and Eve's ancillas ($E$), and $\rho^{BE}=\mathrm{tr_{A}}(\rho^{ABE})$.
Note that system $B$ is released by Bob, encoded by Alice, and returned to Eve.
This PA formula reflects that Eve, in order to learn about the key bit, uses the entire
qubit from Alice and her own ancillas which include those used in her attack on line B-to-A.
This is consistent with the modified protocol where Eve measures $\rho^{BE}$.
We now show that $\rho^{ABE|\text{basis}}=\rho^{ABE}$, i.e., the overall state is not changed
due to the knowledge of the basis.

Bob initially prepares the qubit $\rho^{B}$ in one of the four states.
When the basis is $z$ or $x$, Bob prepares
$\rho^{B|z}=\frac{1}{2}|0\rangle\langle0|+\frac{1}{2}|1\rangle\langle1|$ or $\rho^{B|x}=\frac{1}{2}|+\rangle\langle+|+\frac{1}{2}|-\rangle\langle-|$,
respectively.
Since $\rho^{B|z}=\rho^{B|x}=I/2$, further evolution of this state will
be independent of the basis.
Thus,
\begin{equation}
\label{eqn-rho-ABE1}
\rho^{ABE|\text{basis}}=\rho^{ABE}.
\end{equation}
For completeness, we describe $\rho^{ABE}$ in more detailed as follows.
Eve's most general attack
on B-to-A channel can be described as a joint unitary operation together with
ancilla. After Eve's attack on B-to-A channel, the joint state of the forward
qubit and Eve's ancilla becomes
$\rho^{BE}_{B\rightarrow A}=U_{BE}(\rho^{B}\otimes|E\rangle\langle E|)U_{BE}$,
where $|E\rangle$ is the initial state of Eve's ancilla.
After Alice's encoding operations, the joint state of backward qubit and Eve's ancilla
becomes $\rho^{ABE}=\frac{1}{2}|0\rangle\langle0|^{A}\otimes\rho^{BE}_{0}+
\frac{1}{2}|1\rangle\langle1|^{A}\otimes\rho^{BE}_{1}$, where
$\rho_{0}^{BE}=\rho^{BE}_{B\rightarrow A}$ and
$\rho_{1}^{BE}=Y_{B}\rho^{BE}_{B\rightarrow A}Y_{B}$.

Therefore, Eq.~\eqref{eqn-rho-ABE1} reduces $r_{PA}$ to
$r_{PA} =
S(\rho^{ABE})-S(\rho^{BE})$ for the modified protocol; and also $\rho^{ABE}$
in the modified protocol is the same as in the original protocol because $\rho^{B}$
undergoes the same evolution in both protocols.
Therefore, we can directly apply the calculation of $r_{PA}$ of the original protocol
and its result to the modified protocol here.
In summary, the modified protocol has the same key rate formula Eq.~\eqref{key:rate}.
The four-state protocol is secure independent of the measurement and
is immune against all detector side channel attacks in line A-to-B.

\begin{table}
\begin{tabular}{|c|c|c|}
\hline
& \multicolumn{2}{c|}{Bit value} \\
\cline{2-3}
Basis & 0 & 1\\
\hline
$x$ & $\{I,X\}$ & $\{Z,Y\}$
\\
\hline
$z$ & $\{I,Z\}$ & $\{X,Y\}$
\\
\hline
\end{tabular}
\caption{Key bit value dependence on the basis used by Bob ($x$ or $z$) and Alice's encoding operation ($I$, $X$, $Y$, or $Z$).
For instance, when Bob uses basis $x$, bit $1$ is encoded by Alice if she applies $Z$ or $Y$ on the qubit emitted by Bob.
\label{table-encoded-value}
}
\end{table}

\subsection{Proof of MDI security of the four-state protocol with four encoding operations}
Here, we also show the MDI security in line A-to-B of the four-state protocol with four encoding
operations of Ref.~\cite{fmcc12} [see Fig.~\ref{fig:protocols}(b)].
In this protocol, the steps are the same as the protocol given
in Sec.~\ref{sec-security of the four-state protocol} except with steps (4) and (5) replaced as follows:
(4) In the encoding mode, Alice randomly perform unitary operations
$I=|0\rangle\langle0|+|1\rangle\langle1|$, $X=|1\rangle\langle0|+|0\rangle\langle1|$,
$Z=|-\rangle\langle+|+|+\rangle\langle-|$,
 or $Y=|0\rangle\langle1|-|1\rangle\langle0|$. The actual key bit value is dependent on the basis used
 (Table~\ref{table-encoded-value}).
(5) Alice sends the encoded
qubits back to Bob.
Bob
measures each qubit in the same basis as the one he used for preparation to
get Alice's key bit deterministically, without basis reconciliation.
Bob also tells Alice his basis choice so that she knows the key bit value according to Table~\ref{table-encoded-value}.

We show that this protocol is MDI secure in line A-to-B.
The security proof of this protocol given in Ref.~\cite{fmcc12} reduces the protocol
to an equivalent protocol where the forward channel B-to-A runs the BB84 protocol
and the backward channel A-to-B runs a classical one-time pad (OTP) [see Fig.~\ref{fig:protocols}(c) and Fig.~3 of Ref.~\cite{fmcc12}].
It is argued in Ref.~\cite{fmcc12} that the classical OTP is equivalent to a quantum OTP by encoding
classical states onto quantum states in some basis.
The basis used has no bearing on the security since the security originates from that
of the classical OTP: Eve is allowed to know the OTP-encrypted bit value in transit on line A-to-B.
It is argued further in Ref.~\cite{fmcc12} that this basis is the basis used by Bob for his initial state.
Thus, it is irrelevant whether Bob announces the basis before or after the OTP operation on line A-to-B.
Using the same modification in Sec.~\ref{sec-proof-MDI1} for moving the measurement device from Bob to Eve,
this variant of the four-state protocol is MDI secure in line A-to-B.

\section{discussion and conclusion}

Although the two-way DQKD has higher resistance to PNS attack-like attacks \cite{lm10},
decoy states are necessary to modify the four-state protocol for long-distance key distribution \cite{mk12}.
Another problem is that the two-way QKD is vulnerable to Eve's Trojan horse attacks, such as the invisible
photon attack \cite{cai06}. In practice, Alice can add a filter to defeat Eve's invisible photon attack.
All Bob's photon pulses should pass through Alice's filter first and only wavelengths close to the
operating wavelength are allowed to pass through.

It has been widely believed that, since Eve can attack the travel photons both in line B-to-A and in line
A-to-B, QKD with a two-way quantum channel is more vulnerable under Eve's practical attacks
comparing with one-way QKD protocol, such as the BB84 protocol. In this paper, we have shown that two four-state
protocols are MDI secure in line A-to-B, i.e., when Eve instead of Bob measures Alice's qubit using Bob's basis,
the modified protocols have the same key generation rate as that of the original protocols.
Since the line A-to-B is MDI secure, measurement device dependent attacks may only work in line B-to-A.
Finally, we want to emphasize that we only proved the security of two-way DQKD against detector side
channel attacks at Bob's side in the backward line A-to-B, while the MDI security of Alice's detectors in the check mode
should be studied in future work.

\section*{Acknowledgments}

Financial support from NSFC under Grant Nos. 11104324
and 11074283 of the PRC Government and RGC under Grant 700712P of the HKSAR Government is gratefully acknowledged.

\end{document}